\documentclass[aps,prl,twocolumn,showpacs,superscriptaddress]{revtex4}
\usepackage{epsfig}
\usepackage{graphicx}
\usepackage{color}

\def\beq{\begin{equation}}
\def\eeq{\end{equation}}
\def\bea{\begin{eqnarray}}
\def\eea{\end{eqnarray}}
\def\nn{\nonumber}

\def\sss{\scriptscriptstyle}

\def\roughly#1{\mathrel{\raise.3ex\hbox
{$#1$\kern-.75em\lower1ex\hbox{$\sim$}}}}

\def\sla#1{\raise.15ex\hbox{$/$}\kern-.57em #1}
\def\bra#1{\left\langle #1\right|}
\def\ket#1{\left| #1\right\rangle}

\def\bd{B_d^0}

\def\btos{{\bar b} \to {\bar s}}
\def\ANPq{{\cal A}^q}

\def\ApNPcom{{\cal A}^{\prime, comb} }
\def\ApNPqph{{\cal A}^{\prime,q} e^{i \Phi'_q}}
\def\ApNPCqph{{\cal A}^{\prime {\sss C}, q} e^{i \Phi_q^{\prime C}}}
\def\ApNPCuph{{\cal A}^{\prime {\sss C}, u} e^{i \Phi_u^{\prime C}}}
\def\ApNPCdph{{\cal A}^{\prime {\sss C}, d} e^{i \Phi_d^{\prime C}}}
\def\pewcp{P_{\sss EW}^{\prime\sss C}}
\def\pewp{P'_{\sss EW}}
\def\ApNPuph{{\cal A}^{\prime,u} e^{i \Phi'_u}}
\def\ApNPdph{{\cal A}^{\prime,d} e^{i \Phi'_d}}
\def\ApNPcomb{{\cal A}^{\prime, comb} e^{i \Phi'}}
\def\btopik{B \to \pi K}

\begin{document}

\preprint{UdeM-GPP-TH-07-156} 

\title{\boldmath Is There Still a $\btopik$ Puzzle?}

\author{Seungwon Baek}
\affiliation{Institute of Physics and Applied Physics, Yonsei University,
Seoul 120-479, Korea}
\author{David London} 
\affiliation{Physique des Particules, Universit\'e de Montr\'eal, \\
C.P. 6128, succ.~centre-ville, Montr\'eal, QC, Canada H3C 3J7}

\date{\today}

\begin{abstract} 
We perform a fit to the 2006 $\btopik$ data and show that there is a
disagreement with the standard model. That is, the $\btopik$ puzzle is
still present. In fact, it has gotten worse than in earlier years.
Assuming that one new-physics (NP) operator dominates, we show that a
good fit is obtained only when the electroweak penguin amplitude is
modified. The NP amplitude must be sizeable, with a large weak phase.
\end{abstract}

\maketitle

There are four $\btopik$ decays -- $B^+ \to \pi^+ K^0$ (designated as
$+0$ below), $B^+ \to \pi^0 K^+$ ($0+$), $\bd \to \pi^- K^+$ ($-+$)
and $\bd \to \pi^0 K^0$ ($00$) -- whose amplitudes are related by an
isospin quadrilateral relation. There are nine measurements of these
processes that can be made: the four branching ratios, the four direct
CP asymmetries $A_{\sss CP}$, and the mixing-induced CP asymmetry
$S_{\sss CP}$ in $\bd\to \pi^0K^0$. Several years ago, these
measurements were in disagreement with the standard model (SM),
leading some authors to posit the existence of a ``$\btopik$ puzzle''
\cite{BFRS}. Subsequently, there was an enormous amount of work
looking at $\btopik$ decays, both within the SM and with new physics
(NP).

One simple way to see this discrepancy is to define the ratios of
branching ratios
\begin{eqnarray}
 {R_n} \equiv \frac12 \, \frac{BR[\bd\to\pi^-K^+]+BR[\bar \bd\to\pi^+K^-]}
        {BR[\bd\to\pi^0K^0]+BR[\bar \bd\to\pi^0 \bar K^0]} ~, \nonumber\\
 {R_c} \equiv 2 \, \frac{BR[B_d^+\to\pi^0 K^+]+BR[B_d^-\to\pi^0 K^-]}
        {BR[B_d^+\to\pi^+ K^0]+BR[B_d^-\to\pi^- \bar K^0]} ~.
\end{eqnarray}
Within the SM, it is predicted that $R_c$ is approximately equal to
$R_n$ \cite{sumrule}. QCD-factorization techniques \cite{BBNS} can be
used to take corrections into account. These yield
\cite{Beneke:2003zv}
\beq
R_n=1.16^{+0.22}_{-0.19} ~,~~ R_c=1.15^{+0.19}_{-0.17} ~.
\eeq
However, the pre-ICHEP04 data was \cite{hurth}
\beq
R_n=0.76^{+0.10}_{-0.10} ~,~~ R_c=1.17^{+0.12}_{-0.12} ~.
\eeq
The values of $R_n$ and $R_c$ revealed a clear discrepancy with the
SM. Even the pre-ICHEP06 data was in disagreement with the SM.

However, the ICHEP06 data is
\beq
R_n=1.00^{+0.07}_{-0.07} ~,~~ R_c=1.10^{+0.07}_{-0.07} ~,
\eeq
which is now in agreement with the SM. This has led some authors to
claim that there is no longer a $\btopik$ puzzle
\cite{hurth,nopuzzle}.

In the present paper, we revisit the question of whether or not there
is a $\btopik$ puzzle. As we will see, the $\btopik$ puzzle is still
present, and is in fact more severe than before. However, it now is
present only in CP-violating asymmetries.  If one looks at only the
measurements of the branching ratios, as above, one will not see it.
We also examine the question of which NP scenarios can account for the
current $\btopik$ data. As such, this paper can be considered an
update of Ref.~\cite{Baeketal}, in which the present authors were
involved.

We begin by writing the four $\btopik$ amplitudes using the
diagrammatic approach. Within this approach \cite{GHLR}, the
amplitudes can be written in terms of six diagrams: the color-favored
and color-suppressed tree amplitudes $T'$ and $C'$, the gluonic
penguin amplitudes $P'_{tc}$ and $P'_{uc}$, and the color-favored and
color-suppressed electroweak penguin amplitudes $\pewp$ and
$\pewcp$. (The primes on the amplitudes indicate $\btos$ transitions.)
The amplitudes are given by
\bea
\label{fulldiagrams}
A^{+0} &\!\!=\!\!& -P'_{tc} + P'_{uc} e^{i\gamma} -\frac13
\pewcp ~, \nn\\
\sqrt{2} A^{0+} &\!\!=\!\!& -T' e^{i\gamma} -C' e^{i\gamma}
+P'_{tc} \nn\\ 
& & ~~~~ -~P'_{uc} e^{i\gamma} -~\pewp -\frac23 \pewcp ~, \nn\\
A^{-+} &\!\!=\!\!& -T' e^{i\gamma} + P'_{tc} -P'_{uc}
e^{i\gamma} -\frac23 \pewcp ~, \nn\\
\sqrt{2} A^{00} &\!\!=\!\!& -C' e^{i\gamma} - P'_{tc} +P'_{uc}
e^{i\gamma} - \pewp -\frac13 \pewcp ~,
\eea
where we have explicitly written the weak-phase dependence, (including
the minus sign from $V_{tb}^* V_{ts}$ [$P'_{tc}$]), while the diagrams
contain strong phases. (The phase information in the
Cabibbo-Kobayashi-Maskawa quark mixing matrix is conventionally
parametrized in terms of the unitarity triangle, in which the interior
(CP-violating) angles are known as $\alpha$, $\beta$ and $\gamma$
\cite{pdg}.) The amplitudes for the CP-conjugate processes can be
obtained from the above by changing the sign of the weak phase
($\gamma$).

Within the SM, to a good approximation, the diagrams $\pewp$ and
$\pewcp$ can be related to $T'$ and $C'$ using flavor SU(3) symmetry
\cite{EWPs}:
\bea
\label{EWPrels}
\pewp & \!\!=\!\! & {3\over 4} {c_9 + c_{10} \over c_1 + c_2} R (T' +
C') \!+\!  {3\over 4} {c_9 - c_{10} \over c_1 - c_2} R (T' - C')
~, \nn\\
\pewcp & \!\!=\!\! & {3\over 4} {c_9 + c_{10} \over c_1 + c_2} R (T' +
C') \!-\!  {3\over 4} {c_9 - c_{10} \over c_1 - c_2} R (T' - C')
~.
\eea
Here, the $c_i$ are Wilson coefficients \cite{BuraseffH} and $R \equiv
\left\vert (V_{tb}^* V_{ts})/(V_{ub}^* V_{us}) \right\vert$.  In our
fits we take $R=48.9 \pm 1.6$ \cite{CKMfitter}.

In Ref.~\cite{GHLR}, the relative sizes of the $\btopik$ diagrams
were roughly estimated as
\bea
& 1 : |P'_{tc}| ~~,~~~~ {\cal O}({\bar\lambda}) : |T'|,~|\pewp| ~, & \nn\\
& {\cal O}({\bar\lambda}^2) : |C'|,~|P'_{uc}|,~|\pewcp| ~, &
\label{ampsizes}
\eea
where ${\bar\lambda} \sim 0.2$. These estimates are expected to hold
approximately in the SM. If one ignores the small ${\cal
O}({\bar\lambda}^2)$ diagrams, the four $\btopik$ amplitudes take the
form
\bea
\label{SMamps}
A^{+0} & = & -P'_{tc} ~, \nn\\
\sqrt{2} A^{0+} & = & -T' e^{i\gamma} +P'_{tc} - \pewp , \nn\\
A^{-+} & = & -T' e^{i\gamma} +P'_{tc} ~, \nn\\
\sqrt{2} A^{00} & = & -P'_{tc} - \pewp ~.
\eea

\begin{table}[tbh]
\center
\begin{tabular}{cccc}
\hline
\hline
Mode & $BR[10^{-6}]$ & $A_{\sss CP}$ & $S_{\sss CP}$ \\ \hline
$B^+ \to \pi^+ K^0$ & $23.1 \pm 1.0$ & $0.009 \pm 0.025$ & \\
$B^+ \to \pi^0 K^+$ & $12.8 \pm 0.6$ & $0.047 \pm 0.026$ & \\
$\bd \to \pi^- K^+$ & $19.7 \pm 0.6$ & $-0.093 \pm 0.015$ & \\
$\bd \to \pi^0 K^0$ & $10.0 \pm 0.6$ & $-0.12 \pm 0.11$ &
$0.33 \pm 0.21$ \\
\hline
\hline
\end{tabular}
\caption{Branching ratios, direct CP asymmetries $A_{\sss CP}$, and
mixing-induced CP asymmetry $S_{\sss CP}$ (if applicable) for the four
$\btopik$ decay modes. The data is taken from Refs.~\cite{HFAG} and
\cite{piKrefs}.}
\label{tab:data}
\end{table}

From the above, we see that the amplitudes $\sqrt{2} A(B^+ \to \pi^0
K^+)$ and $A(\bd \to \pi^- K^+)$ are equal, up to a factor of
$\pewp$. Since $|\pewp| < |P'_{tc}|$, we therefore expect the direct
asymmetry $A_{\sss CP}(0+)$ to be approximately equal to $A_{\sss
CP}(-+)$:
\beq
 A_{\sss CP}(0+) \approx
 -2  \left|T' \over P'_{tc} \right| \sin \delta_{\sss T'} \sin \gamma
\approx A_{\sss CP}(-+) ~,
\eeq
where $\delta_{\sss T'}$ is the strong-phase difference between $T'$ and
$P'_{tc}$. However, looking at the ICHEP06 $\btopik$ data in Table 1,
we see that these asymmetries are very different. Thus, the 2006
$\btopik$ data cannot be explained by the SM. It is only by
considering the CP-violating asymmetries that one realizes this.

To investigate this numerically, we have performed a fit to the 2006
ICHEP06 $\btopik$ data using the parametrization of
Eq.~(\ref{SMamps}). We find that $\chi^2_{min}/d.o.f. =
25.0/5~(1.4\times 10^{-4})$, indicating an extremely poor fit. (The
number in parentheses indicates the quality of the fit, and depends on
$\chi^2_{min}$ and $d.o.f.$ individually. 50\% or more is a good fit;
fits which are substantially less than 50\% are poorer.)

The only way to account for the data within the SM is to include $C'$,
since it appears in only one of the two amplitudes $\sqrt{2} A(B^+ \to
\pi^0 K^+)$ and $A(\bd \to \pi^- K^+)$. Thus, as a first step to
remedy the problem, we follow Refs.~\cite{piKfit} and add only $C'$
[from Eq.~(\ref{fulldiagrams})] to the amplitudes of
Eq.~(\ref{SMamps}). From the approximate formulae
\bea
 A_{\sss CP}(0+) &\!\! \approx \!\!&
 -2  \left|T' \over P'_{tc} \right| \sin \delta_{\sss T'} \sin \gamma \\
 A_{\sss CP}(-+) & \!\!\approx \!\!&
 -2  \left|T' \over P'_{tc} \right| \sin \delta_{\sss T'} \sin \gamma
- 2  \left|C' \over P'_{tc} \right| \sin \delta_{\sss C'} \sin \gamma ~, \nn
\eea
where $\delta_{\sss C'}$ is the strong-phase difference between $C'$
and $P'_{tc}$, we see that a large value of $|C'|$ can give the
correct sign for $A_{\sss CP}(-+)$ when $ \sin \delta_{\sss C'}$ has a
different sign from $ \sin \delta_{\sss T'}$. This is confirmed
numerically. A good fit is obtained:
$\chi^2_{min}/d.o.f. =1.0/3~(80\%)$. We find $|P'| = 47 \pm 1$ eV,
$|T'| = 8.1 \pm 3.5$ eV, $|C'| = 13.0 \pm 3.2$ eV, $\delta_{\sss T'} =
(154 \pm 10)^\circ$, $\delta_{\sss C'} = (-154 \pm 7)^\circ$. However,
note that $|C'/T'| = 1.6 \pm 0.3$ is required (we stress that
correlations have been taken into account in obtaining this ratio).
Since this value is much larger than the naive estimates of
Eq.~(\ref{ampsizes}), this shows explicitly that the $\btopik$ puzzle
is still present. In Ref.~\cite{Baeketal} (2004), $|C'/T'| = 1.8 \pm
1.0$ was found. We thus see that the puzzle has gotten much worse in
2006. (See Ref.~\cite{zhou} for another approach.)

We note in passing that the two fits give $\gamma = (62.5 \pm
11.1)^\circ$ and $\gamma = (50.0 \pm 5.6)^\circ$, respectively.  Thus,
both fits give values for $\gamma$ which are consistent with the value
obtained via a fit to independent measurements: $\gamma =
{59.0^{+6.4}_{-4.9}}^\circ$ \cite{CKMfitter}.

We can also perform a fit which incorporates all the ${\cal
O}({\bar\lambda}^2)$ terms in Eq.~(\ref{fulldiagrams}) (i.e.\
$P'_{uc}$ is included). Once again, we find a good fit:
$\chi^2_{min}/d.o.f. = 0.7/1~(79\%)$ (the values of $A_{\sss CP}(0+)$
and $A_{\sss CP}(-+)$ are reproduced due to $C'$--$P'_{uc}$
interference). However, a smaller value of $|C'|$ is found,
compensated by a large value of $|P'_{uc}|$: $|C'/T'|=0.8 \pm 0.1$,
$|P'_{uc}/T'|= 1.7 \pm 0.6$. We also find that $\gamma = (30 \pm
7)^\circ$.  Thus, the problem in $|C'/T'|$ has been alleviated, but we
encounter new difficulties in $|P'_{uc}/T'|$ and $\gamma$.

Having established that there still is a $\btopik$ puzzle, we now turn
to the question of the type of new physics which can be responsible.
There are a great many NP operators which can contribute to $\btopik$
decays. However, this number can be reduced considerably. All NP
operators in $\btos q {\bar q}$ transitions take the form ${\cal
O}_{\sss NP}^{ij,q} \sim {\bar s} \Gamma_i b \, {\bar q} \Gamma_j q$
($q = u,d,s,c$), where the $\Gamma_{i,j}$ represent Lorentz
structures, and color indices are suppressed. These operators
contribute to the decay $\btopik$ through the matrix elements
$\bra{\pi K} {\cal O}_{\sss NP}^{ij,q} \ket{B}$, whose magnitude is
taken to be roughly the same size as the SM $\btos$ penguin operators.
Each matrix element has its own NP weak and strong phase.

We begin with the model-independent analysis described in
Ref.~\cite{DLNP} and used in Ref.~\cite{Baeketal}. Here it is argued
that all NP strong phases are negligible. This allows for a great
simplification. If one neglects all NP strong phases, one can now
combine all NP matrix elements into a single NP amplitude, with a
single weak phase:
\beq
\sum \bra{\pi K} {\cal O}_{\sss NP}^{ij,q} \ket{B} = \ANPq e^{i
\Phi_q} ~.
\eeq
Now, $\btopik$ decays involve only NP parameters related to the quarks
$u$ and $d$. These operators come in two classes, differing in their
color structure: ${\bar s}_\alpha \Gamma_i b_\alpha \, {\bar q}_\beta
\Gamma_j q_\beta$ and ${\bar s}_\alpha \Gamma_i b_\beta \, {\bar
q}_\beta \Gamma_j q_\alpha$ ($q=u,d$). The matrix elements of these
operators can be combined into single NP amplitudes, denoted
$\ApNPqph$ and $\ApNPCqph$, respectively \cite{BNPmethods}. Here,
$\Phi'_q$ and $\Phi_q^{\prime {\sss C}}$ are the NP weak phases; the
strong phases are zero. Each of these contributes differently to the
various $\btopik$ decays. In general, ${\cal A}^{\prime,q} \ne {\cal
A}^{\prime {\sss C}, q}$ and $\Phi'_q \ne \Phi_q^{\prime {\sss
C}}$. Note that, despite the ``color-suppressed'' index $C$, the
matrix elements $\ApNPCqph$ are not necessarily smaller than the
$\ApNPqph$.

The $\btopik$ amplitudes can now be written in terms of the SM
amplitudes to $O({\bar\lambda})$ [$\pewp$ and $T'$ are related as in
Eq.~(\ref{EWPrels})], along with the NP matrix elements
\cite{BNPmethods}:
\bea
\label{BpiKNPamps}
A^{+0} &\!\!=\!\!& -P'_{tc} + \ApNPCdph ~, \\
\sqrt{2} A^{0+} &\!\!=\!\!& P'_{tc} - T' \, e^{i\gamma} -
\pewp \nn\\
& & ~~~~ +~\ApNPcomb - \ApNPCuph ~, \nn\\
A^{-+} &\!\!=\!\!& P'_{tc} - T' \, e^{i\gamma} - \ApNPCuph
~, \nn\\
\sqrt{2} A^{00} &\!\!=\!\!& -P'_{tc} - \pewp
+ \ApNPcomb + \ApNPCdph ~, \nn
\eea
where $\ApNPcomb \equiv - \ApNPuph + \ApNPdph$. 

Note that the above form of NP amplitudes still satisfies the isospin
quadrilateral relation. If $P'_{tc}$ is dominant, one will also find
that $R_n \approx R_c$ and the sum rules for CP asymmetries given in
Ref.~\cite{Gronau_sumrule} are satisfied. Therefore the fact that the
SM satisfies those sum rules does not necessarily rule out this kind
of NP. (Of course, if the NP is about the same size as $P'_{tc}$, one
can find that $R_n \ne R_c$ and that the sum rules are violatied.)

The value of $\gamma$ is taken from independent measurements ($\gamma
= {59.0^{+6.4}_{-4.9}}^\circ$ \cite{CKMfitter}), leaving a total of 10
theoretical parameters: $|P'_{tc}|$, $|T'|$, $|{\cal A}^{\prime,
comb}|$, $|{\cal A}^{\prime {\sss C}, u}|$, $|{\cal A}^{\prime {\sss
C}, d}|$, 3 NP weak phases and two relative strong phases. With only 9
experimental measurements, it is not possible to perform a fit. It is
necessary to make some theoretical assumptions.

As in Ref.~\cite{Baeketal}, we assume that a single NP amplitude
dominates. We consider the following four possibilities: (i) only
${\cal A}^{\prime, comb} \ne 0$, (ii) only ${\cal A}^{\prime {\sss C},
u} \ne 0$, (iii) only ${\cal A}^{\prime {\sss C}, d} \ne 0$, (iv)
$\ApNPCuph = \ApNPCdph$, ${\cal A}^{\prime, comb} = 0$
(isospin-conserving NP).

However, before presenting the results of the fits, we can deduce what
the results should yield. Earlier. we saw that the SM can reproduce
the ICHEP06 $\btopik$ data, but only if $C'$ is anomalously large. Of
the three NP matrix elements, only $\ApNPcomb$ appears only in
amplitudes of decays which receive contributions from $C'$ ($A^{0+}$
and $A^{00} $); $\ApNPCuph$ and $\ApNPCdph$ appear also in amplitudes
with no $C'$ ($A^{+0}$ and $A^{-+} $). Thus, if one wishes to
reproduce a large $C'$ (or $\pewp$ \cite{patterns}), it is necessary
to add $\ApNPcomb$; $\ApNPCuph$ and $\ApNPCdph$ are not necessary. In
light of this, we expect a good fit only for possibility (i) above;
(ii), (iii) and (iv) should have poor fits. (Note that fit (ii) might
be marginally acceptable due to the interference of $\pewp$ and
$\ApNPCuph$.)

This is borne out numerically. We find (i) $\chi^2_{min}/d.o.f. =
0.6/3~(90\%)$, (ii) $\chi^2_{min}/d.o.f. = 4.0/3~(26\%)$, (iii)
$\chi^2_{min}/d.o.f. = 21.1/3~(0.01\%)$, (iv) $\chi^2_{min}/d.o.f. =
12.8/4~(1.2\%)$. (Regarding the $d.o.f.$'s: in fits (i)-(iii), we fit
to the three magnitudes $|P'_{tc}|$, $|T'|$ and $|{\cal A}|$ (NP), two
relative strong phases, and the NP weak phase $\Phi$, so the $d.o.f.$
for these three fits is $9-6=3$. In fit (iv), one of the NP parameters
can be absorbed into $|P'_{tc}|$, so the number of $d.o.f.$'s is
increased by one.) Thus, as expected, a good fit is found only if the
NP is in the form of $\ApNPcomb$.

This is the same conclusion as that found in Ref.~\cite{Baeketal}.
Thus, not only is the $\btopik$ puzzle still present, but it is still
pointing towatds the same type of NP, $\ApNPcomb \ne 0$ (this
corresponds to NP in the electroweak penguin amplitude). For this
(good) fit, we find $|T'/P'| = 0.09$, $|{\cal A}^{\prime, comb}/P'| =
0.24$, $\Phi' = 85^\circ$. We therefore find that the NP amplitude
must be sizeable, with a large weak phase.

If one does not wish to neglect the NP strong phases, one can still
reduce the number of NP operators. It can be shown that an arbitrary
amplitude can be written in terms of two others with known weak
phases. This is known as reparametrization invariance (RI)
\cite{RI_papers}. Applying this to the NP amplitudes, and taking the
two weak phases to be $0$ and $\gamma$, we can write
\beq
\sum \bra{\pi K} {\cal O}_{\sss NP}^{ij,q} \ket{B} = {\cal A}^{q}_0 
+{\cal A}^q_\gamma e^{i \gamma}  ~,
\label{eq:RI}
\eeq
where the ${\cal A}^{q}_{(0,\gamma)}$ contain only strong phases.

The $\btopik$ amplitudes can now be written as in
Eq.~(\ref{BpiKNPamps}), where each of the NP amplitudes is expressed
as in Eq.~(\ref{eq:RI}) above. Taking $\gamma =
{59.0^{+9.2}_{-3.7}}^\circ$ \cite{CKMfitter}, there are a total of 15
theoretical parameters: the eight magnitudes $|P'_{tc}|$, $|T'|$,
$|{\cal A}^{\prime, comb}_{(0,\gamma)}|$, $|{\cal A}^{\prime {\sss C},
u}_{(0,\gamma)}|$, $|{\cal A}^{\prime {\sss C}, d}_{(0,\gamma)}|$, and
seven relative strong phases. With only 9 measurements, we must again
make theoretical assumptions, and we consider the four possibilities
(i)-(iv) described above.

As before, we expect a good fit only for possibility (i). For the
$d.o.f.$'s, in the fits there are seven theoretical parameters: the
four magnitudes $|P'_{tc}|$, $|T'|$, $|{\cal A}^{q}_0|$ and $|{\cal
A}^{q}_\gamma|$, and three relative strong phases. In fit (ii),
$A_{\sss CP}(+0)$ and $A_{\sss CP}(00)$ are identically zero, and
$S_{\sss CP}(00) = \sin 2\beta$. Thus, these measurements do not
constrain the parameters; there are only six constraining
measurements. This is fewer than the number of parameters, so a fit
cannot be done in this case.  The fit can be done in cases (i), (iii),
and the $d.o.f.$ is $9-7=2$.  In fit (iv), ${\cal A}^{q}_0$ can be
absorbed into $P'_{tc}$, so the $d.o.f.$ is $9-5=4$. The results of
the fits are: (i) $\chi^2_{min}/d.o.f. = 0.5/2~(79\%)$, (iii)
$\chi^2_{min}/d.o.f. = 18.5/2~(0.01\%)$, (iv) $\chi^2_{min}/d.o.f. =
12.8/4~(1.2\%)$. As expected, only fit (i) [NP $=\ApNPcom$] gives a
good fit. We find $|T'/P'| = 0.09$, $|{\cal A}^{q}_0/P'| = 0.15$,
$|{\cal A}^{q}_0/P'| = 0.28$. We therefore find that the NP amplitude
must be sizeable.

We have considered two different parametrizations of the NP effects:
negligible NP strong phases, and the general case. Although the
results of the fits are not numerically identical, the conclusions are
similar.

Of the four new-physics models examined in this paper, only one
produces a good fit to the 2006 $\btopik$ data: case (i), ${\cal
A}^{\prime, comb} \ne 0$. The quality of fit is extremely good
here. The NP amplitude must be sizeable, with a large weak phase. Fit
(ii) (${\cal A}^{\prime {\sss C}, u} \ne 0$) is marginal if the NP
strong phases are negligible (the fit cannot be done in the general
case). Here $A_{\sss CP}(0+) \approx A_{\sss CP}(-+)$ is found due to
the interference of $\pewp$ and $\ApNPCuph$.  Fits (iii) (${\cal
A}^{\prime {\sss C}, d} \ne 0$) and (iv) (isospin-conserving NP) yield
very poor fits, and are ruled out.

As mentioned in Ref.~\cite{Baeketal}, we note that we have assumed
that one NP operator dominates. However, any specific NP model will
generally have more than one effective NP operator, and so the more
general case might be used to explain the $\btopik$ data.

In conclusion, in this paper we have done a fit to the 2006 $\btopik$
data to see if new physics (NP) was still indicated. Note that a fit
to the full data is necessary; if a subset of the measurements is
used, erroneous conclusions can be reached. If only the ``large''
diagrams of the SM are included in the amplitudes, a very poor fit is
found. If the ``small'' diagram $C'$ is added, a good fit is
obtained. However, $|C'/T'| = 1.6 \pm 0.3$ is required. The fact that
this ratio is large indicates that the $\btopik$ puzzle is still
present. The fact that the error is so small shows that the puzzle has
gotten much worse in 2006.

We have also examined which type of NP can be responsible for the
$\btopik$ puzzle. We assume that one NP operator dominates, and
consider (i) only ${\cal A}^{\prime, comb} \ne 0$, (ii) only ${\cal
A}^{\prime {\sss C}, u} \ne 0$, (iii) only ${\cal A}^{\prime {\sss C},
d} \ne 0$, (iv) isospin-conserving NP: $\ApNPCuph = \ApNPCdph$, ${\cal
A}^{\prime, comb} = 0$. Of these, the only (very) good fit is found
for model (i). It corresponds to a modification of the electroweak
penguin amplitude. Note that these results are the same as those found
in Ref.~\cite{Baeketal}. This shows that, although the puzzle is worse
than in 2004, its resolution is the same.

\bigskip
\noindent
{\bf Acknowledgments}:
This work was financially supported by Brain Korea 21 (BK21) project
(SB) and by NSERC of Canada (DL).



\begin{thebibliography}{99}

\bibitem{BFRS} A.~J.~Buras, R.~Fleischer, S.~Recksiegel and F.~Schwab,
Phys.\ Rev.\ Lett.\ {\bf 92}, 101804 (2004), Nucl.\ Phys.\ B {\bf
697}, 133 (2004), PoS {\bf HEP2005}, 193 (2006).

\bibitem{sumrule} H.~J.~Lipkin, Phys.\ Lett.\ B {\bf 445} (1999) 403;
M.~Gronau and J.~L.~Rosner, Phys.\ Rev.\ D {\bf 59} (1999) 113002;
J.~Matias, Phys.\ Lett.\ B {\bf 520} (2001) 131.

\bibitem{BBNS} M.~Beneke, G.~Buchalla, M.~Neubert and C.~T.~Sachrajda,
Phys.\ Rev.\ Lett.\ {\bf 83}, 1914 (1999), Nucl.\ Phys.\ B {\bf 591},
313 (2000), Nucl.\ Phys.\ B {\bf 606}, 245 (2001).

\bibitem{Beneke:2003zv} M.~Beneke and M.~Neubert, Nucl.\ Phys.\ B {\bf
  675} (2003) 333.

\bibitem{hurth} T.~Hurth, arXiv:hep-ph/0612231.

\bibitem{nopuzzle} For example, see F. Schwab, talk given in
  http://euroflavour06.ifae.es.

\bibitem{Baeketal} S.~Baek, P.~Hamel, D.~London, A.~Datta and
D.~A.~Suprun, Phys.\ Rev.\ D {\bf 71}, 057502 (2005). 

\bibitem{GHLR} M.~Gronau, O.~F.~Hernandez, D.~London and J.~L.~Rosner,
Phys.\ Rev.\ D {\bf 50}, 4529 (1994), Phys.\ Rev.\ D {\bf 52}, 6374
(1995).

\bibitem{pdg} W.~M.~Yao {\it et al.}  [Particle Data Group], J.\
Phys.\ G {\bf 33} (2006) 1.

\bibitem{EWPs} M.~Neubert and J.~L.~Rosner, Phys.\ Lett.\ B {\bf 441},
403 (1998), Phys.\ Rev.\ Lett.\ {\bf 81}, 5076 (1998); M.~Gronau,
D.~Pirjol and T.~M.~Yan, Phys.\ Rev.\ D {\bf 60}, 034021 (1999)
[Erratum-ibid.\ D {\bf 69}, 119901 (2004)]; M.~Imbeault,
A.~L.~Lemerle, V.~Page and D.~London, Phys.\ Rev.\ Lett.\ {\bf 92},
081801 (2004).

\bibitem{BuraseffH} See, for example, G. Buchalla, A.J. Buras and
  M.E. Lautenbacher, {\it Rev.\ Mod.\ Phys.} {\bf 68}, 1125 (1996).

\bibitem{CKMfitter} CKMfitter Group, J.~Charles {\it et al.}, Eur.\
Phys.\ J.\ C {\bf 41}, 1 (2005).

\bibitem{HFAG} Heavy Flavor Averaging Group (HFAG),
arXiv:hep-ex/0603003.

\bibitem{piKrefs} W.~M.~Yao {\it et al.}  [Particle Data Group],
  Ref.~\cite{pdg}; B.~Aubert {\it et al.}  [BABAR Collaboration],
  Phys.\ Rev.\ Lett.\ {\bf 97}, 171805 (2006); K.~Abe {\it et al.}
  [Belle Collaboration], arXiv:hep-ex/0608049; CLEO Collaboration,
  A.~Bornheim {\it et al.}, Phys.\ Rev.\ D {\bf 68}, 052002 (2003);
  B.~Aubert {\it et al.}  [BABAR Collaboration], arXiv:hep-ex/0607106;
  B.~Aubert {\it et al.}  [BABAR Collaboration], rXiv:hep-ex/0608003;
  B.~Aubert {\it et al.}  [BABAR Collaboration], arXiv:hep-ex/0607096;
  S.~Chen {\it et al.}  [CLEO Collaboration], Phys.\ Rev.\ Lett.\ {\bf
  85}, 525 (2000); K.~Abe {\it et al.}  [Belle Collaboration],
  arXiv:hep-ex/0609006.

\bibitem{piKfit} C.~W.~Chiang, M.~Gronau, J.~L.~Rosner and
D.~A.~Suprun, Phys.\ Rev.\ D {\bf 70}, 034020 (2004); S.~Baek, JHEP
{\bf 0607}, 025 (2006).

\bibitem{zhou} Y.~L.~Wu and Y.~F.~Zhou, Phys.\ Rev.\ D {\bf 72} (2005)
  034037; C.~W.~Chiang and Y.~F.~Zhou, JHEP {\bf 0612}, 027 (2006);
  Y.~L.~Wu, Y.~F.~Zhou and C.~Zhuang, Phys.\ Rev.\ D {\bf 74}, 094007
  (2006).

\bibitem{DLNP} A.~Datta and D.~London, Phys.\ Lett.\ B {\bf 595}, 453
(2004).

\bibitem{BNPmethods} A.~Datta, M.~Imbeault, D.~London, V.~Page,
N.~Sinha and R.~Sinha, hep-ph/0406192.

\bibitem{Gronau_sumrule} M.~Gronau and J.~L.~Rosner, Phys.\ Rev.\ D
{\bf 74}, 057503 (2006).

\bibitem{patterns} M.~Imbeault, D.~London, C.~Sharma, N.~Sinha and
R.~Sinha, arXiv:hep-ph/0608169.

\bibitem{RI_papers} F.~J.~Botella and J.~P.~Silva, Phys.\ Rev.\ D {\bf
71}, 094008 (2005); S.~Baek, F.~J.~Botella, D.~London and J.~P.~Silva,
Phys.\ Rev.\ D {\bf 72}, 036004 (2005), Phys.\ Rev.\ D {\bf 72},
114007 (2005).

\end{thebibliography}
\end{document}